\documentclass[11pt]{article}
\usepackage{amsfonts}
\usepackage{amsmath}
\usepackage{amssymb}
\usepackage{extarrows}
\usepackage[
colorlinks,
linkcolor = blue,
citecolor = blue,
urlcolor = blue]{hyperref}

\def \qed {\hfill \vrule height7pt width 7pt depth 0pt}
\usepackage{amssymb}
\usepackage{amsmath}
\usepackage{enumerate}
\usepackage{graphicx}
\usepackage{epstopdf}
\usepackage{amsfonts}
\usepackage{url}
\usepackage{bm}
\usepackage{mathdots}
\usepackage{rotating}
\usepackage{pgfplots}
\usepackage{floatrow}
\usepackage{multicol}

\usepackage{pifont}
\usepackage{epsfig,subfigure,dsfont,amsthm,amsbsy,mathrsfs,amscd}
\usepackage{caption}

\begin{document}
	\title{The local  distinguishability  of any three generalized Bell   states }
	\author{Yan-Ling Wang$^{1}$, Mao-Sheng Li$^{2}$, Shao-Ming Fei$^{3,4}$, Zhu-Jun Zheng$^{1}$\\
		{\footnotesize  {$^1$Department of Mathematics,
				South China University of Technology, Guangzhou
				510640, P.R.China}} \\
		{\footnotesize  {$^2$Department of Mathematical of Science,
				Tsinghua  University,  Beijing
				100084, P.R.China}} \\
		{\footnotesize{
				$^3$School of Mathematical Sciences, Capital Normal University,
				Beijing 100048, China}}\\
		{\footnotesize{$^4$Max-Planck-Institute for Mathematics in the Sciences, 04103
				Leipzig, Germany}}
	}
	
	\maketitle

	\begin{abstract}

We study the problem of distinguishing maximally entangled quantum states by using local operations and classical communication (LOCC). A question of fundamental interest is whether any three maximally entangled states in $\mathbb{C}^d\otimes\mathbb{C}^d (d\geq 4)$ are distinguishable by LOCC. In this paper, we restrict ourselves to consider the  generalized Bell states. And we prove that any three generalized Bell states in $\mathbb{C}^d\otimes\mathbb{C}^d (d\geq 4)$ are locally distinguishable.
	\end{abstract}

	\maketitle
	
	\section{Introduction}
	
	Global operators can not be implemented generally by using only local operations and classical communication (LOCC) in compound quantum systems. Hence it is interesting to  understand  the limitation of quantum operators that can be
	implemented by LOCC. The local distinguishability of quantum states plays
	important roles in exploring the ability of LOCC \cite{Bennett99,Walgate02}. Suppose Alice and Bob share an unknown bipartite quantum state chosen from a given specific set of mutually orthogonal states. Their task is to identify the shared state by using LOCC.  Throughout the paper, the words ``locally distinguishable", ``distinguished  with LOCC" and ``locally distinguished" have the same meanings.
	The local distinguishability has also practical applications in quantum cryptography primitives such as  data hiding \cite{DiVincenzo02}.
	
  According to the property of  mutually orthogonal quantum states to be distinguished, the local distinguishability problem  can be classified as three cases: maximally entangled states, product states and general states. In 2000,
	Walgate et. al. showed that any two orthogonal pure states can be locally distinguishable  \cite{Walgate00}.
	It has been observed in \cite{Ghosh01,Ghosh04,Fan04,Nathanson05} that any set of  maximally entangled states in $\mathbb{C}^d\otimes\mathbb{C}^d$ can not be locally distinguished for large $d$. The lower bound of the numbers of maximally entangled states that are not locally distinguishable has been extensively investigated \cite{Duan12,Cosentino13,CosentinoR14,Li15,Yu15}.
	Locally indistinguishable sets of $d$ maximally entangled states in $\mathbb{C}^d\otimes\mathbb{C}^d$ systems are  constructed for all $d\geqslant4$\cite{Li15,Yu15}. Smaller sets of locally indistinguishable maximally entangled states can be found in \cite{CosentinoR14,Yu15}. Due to the difficulty  of the problem, some researchers studied an easier problem: the one-way local distinguishabililty of maximally entangled states \cite{Bandyopadhyay11,Nathanson2013,Zhang14,Zhang15,Wang16}. For the case of $\mathbb{C}^d\otimes\mathbb{C}^d$ with $d\geq4$, a set of $3\lceil\sqrt{d}\rceil-1$
	one-way LOCC indistinguishable maximally entangled states, which are generalized Bell states, has been constructed \cite{Wang16}.
	
	On the other hand, one can consider the upper bound of the number of maximally entangled states that are locally distinguishable. In 2004, Fan \cite{Fan04} showed that
	if $d$ is prime, then any $k$ mutually orthogonal generalized Bell states
	can be locally distinguished if $k(k-1)<2d$. For $d=3$, any three generalized Bell states can be locally distinguished.
	In \cite{Nathanson05}, it has been shown that in $\mathbb{C}^3\otimes\mathbb{C}^3$, any three mutually orthogonal maximally
	entangled states can be distinguished by LOCC. However, their approaches can not be extended to higher dimensional case. Since then it has been an open question whether
	any three mutually orthogonal maximally entangled states in high dimensions
	can be distinguished  with LOCC.  In 2013, Nathanson presented some examples for triples of  maximally entangled states that cannot   be distinguished with one-way LOCC but two-way \cite{Nathanson2013}. Moreover, Nathanson proved that  any three mutually orthogonal maximally entangled states in $\mathbb{C}^d\otimes\mathbb{C}^d$, $d\geq 3$,
	can be   distinguished with a PPT measurement.   In 2015, Tian \textit{et al} extended Fan's result to quantum systems with dimension of prime power by considering the mutually commuting qudit lattice states \cite{Tian15}. And Singal \textit{et al} give a  complete analysis of perfect local distinguishability of four generalized Bell states in $\mathbb{C}^4\otimes\mathbb{C}^4$ \cite{Singal15}. As an open question remained, it is interesting to consider whether any three  mutually orthogonal generalized Bell states can be locally distinguished for an arbitrary  dimension  $d$.
	
	In this paper, 	we mainly restict ourselves to the locally distinguishable of generalized Bell states.   We first give some properties of the generalized Bell states. We first prove an equation by
employing the method in \cite{Fan04}. By using this equation and some annoyed analysis,   we prove the local distinguishability of any three generalized Bell states case by case. We also solve some exceptional cases by showing the strategies of Alice and Bob employed in order to distinguish the given three states.

	\section{Properties of Generalized Bell states}
	
	Throughout the paper, we use the following notations. In the bipartite system $\mathbb{C}^d\otimes\mathbb{C}^d$, under the computational basis $\{|i\rangle\}_{i=0}^{d-1}$, ${|\psi_0\rangle=\frac{1}{\sqrt{d}}\displaystyle\sum_{i=0}^{d-1}|ii\rangle}$ is a canonical maximally entangled state. In general, a maximally entangled state can be written in the form  $|\psi\rangle=(U\otimes I)|\psi_0\rangle$   with   a unitary matrix  $ U  $. The following $d^2$  maximally entangled states are well known as the generalized Bell states:
\begin{equation}\label{eq1}
\{|\psi_{m,n}\rangle=(U_{m,n}\otimes I)|\psi_0\rangle\big| U_{m,n}=X^mZ^n, m,n=0,1,\cdots,d-1\},
\end{equation}
	  where  $X=\displaystyle\sum_{l=0}^{d-1}|l+1 \text{ mod } d\rangle\langle l|$, and  $Z=\displaystyle\sum_{i=0}^{d-1}\omega^i|i\rangle\langle i|$ with $\omega=e^{\frac{2\pi \sqrt{-1} }{d}}$.
	
	We define $d$   operators $H_{\alpha}, \ \alpha=0,1,...,d-1$, with the entries of $H_{\alpha}$ given by $(H_{\alpha})_{jk}=\omega^{-jk-\alpha s_k}$, $j,k=0,1,\cdots,d-1$, where
	$s_k=k+(k+1)+\cdots+(d-1), k=0,1,\cdots,d-1$. In particular,
we set $s_d=s_0=\dfrac{d(d-1)}{2}$. Then   $\frac{1}{\sqrt{d}}H_{\alpha}$ is unitary for every ${\alpha}$.

	
Motivated by the method in \cite{Fan04} for prime dimensions, we first prove a generalized equation.
	
	\bigskip
	\noindent\textbf{Lemma 1.} The following  equation is satisfied up to a whole phase for all $\alpha$ when $d$ is odd and for even $\alpha$ when $d$ is even.
\begin{equation}	
H_{\alpha}X^mZ^nH_{\alpha}^{\dag}=X^{\alpha m+n}Z^{-m}.
\end{equation}
	
	\noindent{\textit{Proof}:} Since $(s_q-s_{q+1}=q$ when $q=0,1,...,d-2$ and $\omega^{\alpha s_d}=\omega^{\alpha s_0}=1$, then $\omega^{\alpha(s_{d-1}-s_d)}=\omega^{\alpha s_{d-1}}=\omega^{\alpha (d-1)}.$  We have $\omega^{\alpha (s_q-s_{q+1}) }=\omega^{\alpha q}$ and
\begin{equation}\label{HXH}	
	\begin{array}{lcl}
	H_{\alpha}XH_{\alpha}^{\dag}&=&
	\displaystyle\sum_{j,k=0}^{d-1}\omega^{-jk-\alpha s_k}|j\rangle\langle k|\cdot\displaystyle\sum_{i=0}^{d-1}|i+1\rangle\langle i|\cdot\displaystyle\sum_{p,q=0}^{d-1}\omega^{pq+\alpha s_q}|q\rangle\langle p|\\
	&=&\displaystyle\sum_{j,k,p,q=0}^{d-1}\omega^{-jk-\alpha s_k+pq+\alpha s_q}|j\rangle\langle k-1|q\rangle\langle p|\\
	&=& \displaystyle\sum_{j,p,q=0}^{d-1}\omega^{-j(q+1)-\alpha s_{q+1}+pq+\alpha s_q}|j\rangle\langle p|\\
	&=& \displaystyle\sum_{j,p,q=0}^{d-1}\omega^{-j+(-j+\alpha+p)q}|j\rangle\langle p|\\
	&=& \displaystyle\sum_{p=0}^{d-1}\omega^{-\alpha-p}|p+\alpha \rangle\langle p|=Z^{-1}X^{\alpha}.
	\end{array}
\end{equation}

\begin{equation}\label{HZH}
	\begin{array}{lcl}
	H_{\alpha}ZH_{\alpha}^{\dag}&=&\displaystyle\sum_{j,k=0}^{d-1}\omega^{-jk-\alpha s_k}|j\rangle\langle k|\cdot\displaystyle\sum_{i=0}^{d-1}\omega^i|i\rangle\langle i|\cdot\displaystyle\sum_{p,q=0}^{d-1}\omega^{pq+\alpha s_q}|q\rangle\langle p|\\
	&=&\displaystyle\sum_{j,k,p,q=0}^{d-1}\omega^{-(j-1)k-\alpha s_k+pq+\alpha s_q}|j\rangle\langle k|q\rangle\langle p|\\
	&=&\displaystyle\sum_{j,p,q=0}^{d-1}\omega^{(p+1-j)q}|j\rangle\langle p|\\
	&=&\displaystyle\sum_{p=0}^{d-1}|p+1\rangle\langle p|=X.
	\end{array}
\end{equation}
	By using   equations (\ref{HXH}) and (\ref{HZH}), it is easy to derive  $H_{\alpha}X^mZ^nH_{\alpha}^{\dag}=X^{\alpha m+n}Z^{-m}$ up to a whole phase. \qed

	Since the local  distinguishablility of a set of quantum states  is unchanged under arbitary local unitary operators. To locally distinguish a set of generalized Bell states, we first
	let Alice and Bob do unitary operations $\frac{1}{\sqrt{d}}H_{\alpha}$ and $(\frac{1}{\sqrt{d}}H_{\alpha})^t$, respectively,
	where $t$ stands for transposition. This operation is equivalent to
	the transformation $\frac{1}{d}H_{\alpha} X^{m_i} Z^{n_i}H_{\alpha}$ on the Alice side. That is,
	
\begin{equation}	(\frac{1}{\sqrt{d}}H_{\alpha}\otimes (\frac{1}{\sqrt{d}}H_{\alpha})^t)(X^{m_i} Z^{n_i}\otimes I)|\psi_0\rangle =\frac{1}{d}H_{\alpha} X^{m_i}
	Z^{n_i}H_{\alpha}\otimes I|\psi_0\rangle.
\end{equation}

 Here the normalization factor $\frac{1}{\sqrt{d}}$ in $\frac{1}{\sqrt{d}}H_{\alpha}$ does not affect the local distinguishability of the quantum states. We will ignore the factor $\frac{1}{\sqrt{d}}$ and just consider $H_{\alpha}$ as a unitary matrix. From Lemma \ref{eq1}, we know that the transformations $\frac{1}{\sqrt{d}}H_{\alpha}\otimes (\frac{1}{\sqrt{d}}H_{\alpha})^t$ transfer the set of generalized Bell states into itself provided that $\alpha$ satisfies the conditions in Lemma \ref{eq1}.

	The following lemma has been mentioned without a proof in ref.\cite{Fan04}. We give an explicit proof here.
	
	\bigskip
	
	\noindent\textbf{Lemma  2.}\label{lemma} A set of generalized Bell states $\{|\psi_{m_i n_i}\rangle=(U_{m_i n_i}\otimes I) |\psi_0\rangle\}_{i=1}^N$ can be   distinguished under LOCC, if $m_i\neq m_j$  for all $ i\neq  j$  or  $n_i\neq n_j$  for all  $i\neq  j.$

	\noindent\textit{ Proof}: If  $n_i\neq n_j$ for all $ i\neq  j$, then we apply a transformation $H_{\alpha}\otimes (H_{\alpha}^\dagger)^t$ on the given states:
\begin{equation}\label{tran}
(H_{\alpha}\otimes (H_{\alpha}^\dagger)^t)(X^{m_i} Z^{n_i}\otimes I)|\psi_0\rangle =H_{\alpha} X^{m_i}
	Z^{n_i}H_{\alpha}^\dagger\otimes I|\psi_0\rangle.
	\end{equation}
	By Lemma \ref{eq1}, we have the following equations:
\begin{equation}\label{tran2}
H_{\alpha}X^{m_i}Z^{n_i}H_{\alpha}^{\dag}=X^{\alpha m_i+n_i}Z^{-m_i}.
 \end{equation}

	Taking $\alpha=0$ in equations (\ref{tran}) and (\ref{tran2}), we have that the transformation $H_{0}\otimes (H_{0}^\dagger)^t$ transforms  $U_{m_i n_i}\otimes I |\psi_0\rangle$ to $U_{m_i^{'} n_i^{'}}\otimes I |\psi_0\rangle$ with  $m_i^{'}\neq m_j^{'}$   for all  $ i\neq  j$ since $m_i^{'}=n_i$ and $m_j^{'}=n_j$. Hence, we only need to consider the former case.
	
	Suppose   $m_i\neq m_j$ for all $ i\neq  j$.  Alice starts by performing a rank-one projective
	measurement corresponding to the following orthonormal basis: $\{|i\rangle\}_{i=0}^{d-1}$. For each outcome of Alice¡¯s measurement, the post measurement set will be of the following form, up to an irrelevant phase:
\begin{equation}
\{|\psi_{m_1 n_1}\rangle,|\psi_{m_2 n_2}\rangle,...,|\psi_{m_N n_N}\rangle\}\longrightarrow\{|k\rangle|k+m_1\rangle,|k\rangle|k+m_2\rangle,...,|k\rangle|k+m_N\rangle\}.
\end{equation}
Then the Bob's reduced states are orthogonal each other. Thus, once Alice tells Bob her measurement outcome $k$, Bob needs to perform measurement in the $\{|j\rangle\}_{j=0}^{d-1}$ basis. If the outcome of Bob's measurement is $k+m_i$, then the state they shared is $|\psi_{m_i,n_i}\rangle.$
	\qed
	
	\textbf{Remark}: Since the local unitrary transformation does not change the local distinguishability of quantum states, any set of states that can be transformed into a set of states satisfying the conditions of Lemma \ref{lemma} is locally distinguishable.
	
	\section{Local distinguishability of three generalized Bell states}
	
	In this section, we use the unitary matrix     $ X^{m_i}Z^{n_i}  $  to represent the maximally entangled state $|\psi_{m_i,n_i}\rangle= X^{m_i}Z^{n_i}\otimes I |\psi_0\rangle$. We call $|\psi_{m_i,n_i}\rangle$  the state corresponding to  $ X^{m_i}Z^{n_i}  $.
		
	\noindent {\bf Theorem. }In $\mathbb{C}^d\otimes\mathbb{C}^d$ with $d\geq4$, any three states in $S=\{|\psi_{m,n}\rangle|m,n=0,1,\cdots,d-1\}$ are locally distinguishable.
	
	\noindent \emph{Proof:}
The proof is based on the  remark of Lemma \ref{lemma}.	We first give two observations to simplify the problem.

\textbf{	\textit{Observation 1:}} We only need to consider the case
	$\{X^{m_1}Z^{n_1} ,$ $X^{m_1}Z^{n_2} ,$ $X^{m_2}Z^{n_2} \}.$
	By Lemma \ref{lemma}, we have shown that the states corresponding to the matrices  $\{X^{m_i}Z^{n_i} \}_{i=1}^3$   are local distinguishable  with different $m_i$ or $n_i$. Hence we can assume $m_1=m_2$ and $n_1\neq n_2$. If $n_1,n_2,n_3$ are all different, then by Lemma \ref{lemma}, the three states can also be locally distinguished. Hence we can assume $n_3 = n_2$ (or equivalently $n_3= n_1$).
	
\textbf{	\textit{Observation 2: }} Accounting to the following transformations which do not change the local distinguishability:
\begin{equation}
	\begin{array}{l}
	(X^{-m_1}\otimes Z^{{-n_2}^t})(X^{m_1}Z^{n_1}\otimes I)|\psi\rangle=(Z^{n_1-n_2}\otimes I) |\psi\rangle,\\
	(X^{-m_1}\otimes Z^{{-n_2}^t})(X^{m_1}Z^{n_2}\otimes I)|\psi\rangle=I\otimes I|\psi\rangle,\\
	(X^{-m_1}\otimes Z^{{-n_2}^t})(X^{m_2}Z^{n_2}\otimes I)|\psi\rangle=(X^{m_2-m_1}\otimes I)|\psi\rangle,
	\end{array}
\end{equation}
we only need to consider the case $\textit{S}_0=\{I , X^m , Z^n \}$ ($0<m,n<d$).

\begin{figure}[h]
	\normalsize
	\includegraphics[width=1.02\textwidth,height=0.7\textwidth]{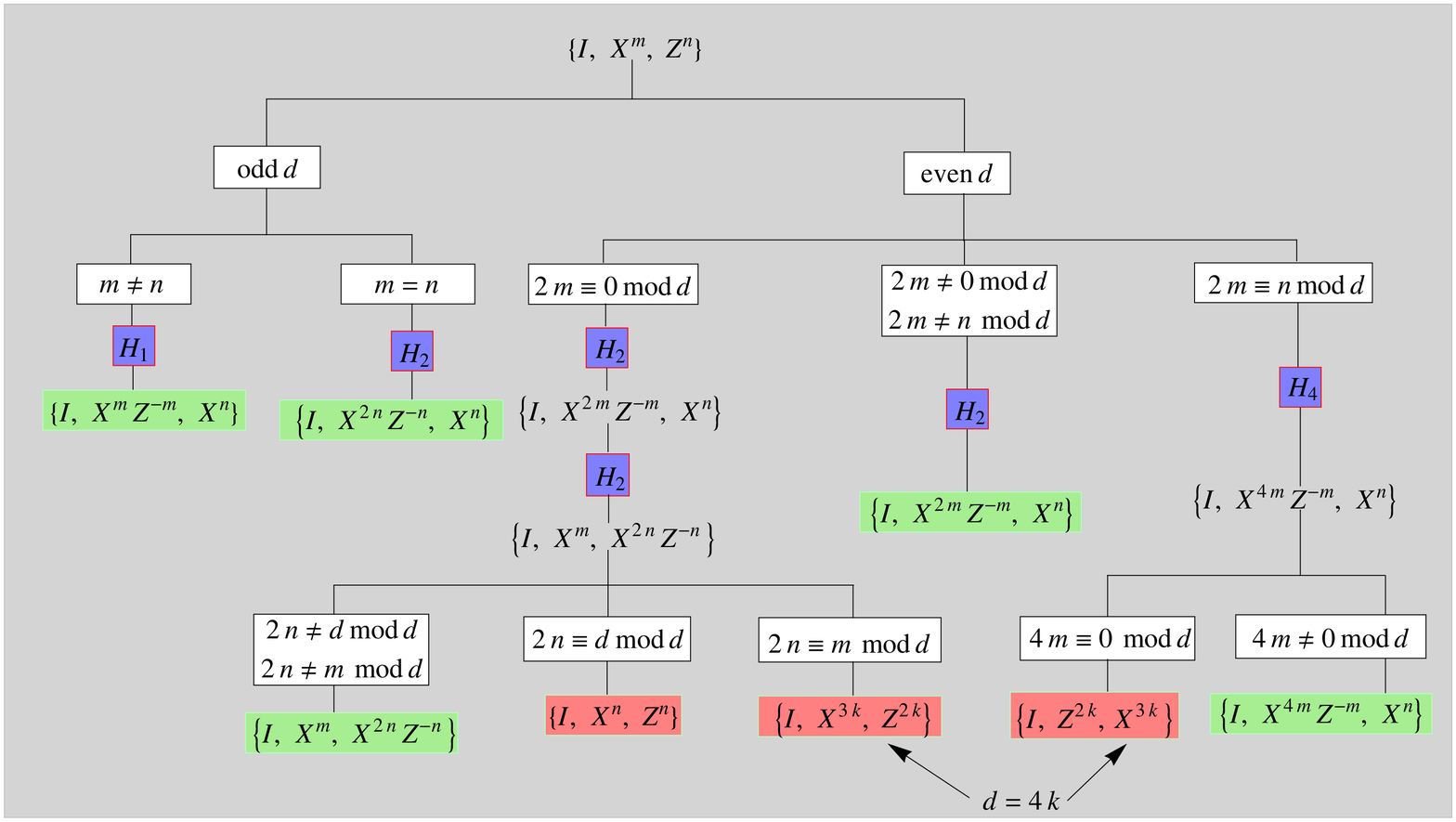}\label{case}
	\caption*{\textbf{Fig. 1 } The figure shows the sketch of the proof. The words in white squares give the conditions. The $H_{\alpha},{\alpha}=1,2,4$, in the blue squares give the transformations. The states with green color can be shown to be locally distinguished by Lemma \ref{lemma}, while those with pink color are the ones called exceptional cases. }
\end{figure}
	
From the above observations, we only need to prove the local distinguishability of the states $\{I , X^m , Z^n \}$.  We use      $X^{m_i} Z^{n_i}  \xlongrightarrow{H_{\alpha} }  H_{\alpha}X^{m_i} Z^{n_i} H_{\alpha}^{\dag} $ to represent  the following transformation (\ref{tr}),
\begin{equation}\label{tr}
(H_{\alpha}\otimes (H_{\alpha}^\dagger)^t)(X^{m_i} Z^{n_i}\otimes I)|\psi_0\rangle =H_{\alpha} X^{m_i}
Z^{n_i}H_{\alpha}^\dagger\otimes I|\psi_0\rangle.	
\end{equation}
Hence, acting $H_\alpha $   on the states $\{I , X^m , Z^n \}$, we obtain $\{I , H_{\alpha}X^mH_{\alpha}^{\dag} ,H_{\alpha}Z^nH_{\alpha}^{\dag} \}.$
	We separate our proof into two cases by the parity of $d$, see Fig. \ref{eq1}.
	
	\textbf{Case I:}
	If $d$ is odd, then for any $\alpha$,  $H_{\alpha}X^mH_{\alpha}^{\dag}=X^{\alpha m}Z^{-m}$, and $H_{\alpha}Z^nH_{\alpha}^{\dagger}=X^n.$ Under the transformation $H_\alpha $, the states $\{I , X^m  , Z^n \}$ are transfered to the states $\{I , X^{\alpha m}Z^{-m} , X^n \}$.

	i) $m\neq n$, $\{I , X^m , Z^n \} \xlongrightarrow{H_{1}  }\textit{S}_1=\{I , X^{m}Z^{-m} , X^n \} $. Note that the states $\{0,m,n$   mod  $ d\}$ are not equal each other. By Lemma \ref{lemma}, the set $\textit{S}_1$ of states are locally distinguishable.
	
	ii) $m=n$, $\{I , X^n , Z^n \} \xlongrightarrow{H_{2} }\textit{S}_2=\{I , X^{2n}Z^{-n} , X^n \}. $ Also it is easy to show that $\{0,2n,n$  mod  $d\}$ are not equal each other for odd $d$. Then by Lemma \ref{lemma}, the set $\textit{S}_2$ of states are locally distinguishable.
	
	\textbf{Case II:}
	If  $d$ is even, $H_{\alpha}X^mH_{\alpha}^{\dag}=X^{\alpha m}Z^{-m}$ and $H_{\alpha}Z^nH_{\alpha}^{\dagger}=X^n$ are satisfied for even  $\alpha$. Consider the following transformation,
\begin{equation}
\{I, X^m, Z^n\} \xlongrightarrow{H_{2} }\textit{S}_2=\{I, X^{2m}Z^{-m}, X^n\}.
\end{equation}

	i) If $2m\neq 0 \text{ mod } d,\text{ and } 2m\neq n \text{ mod } d$, then $\{0,2m,n\text{ mod } d\}$ are not identical. By Lemma \ref{lemma}, the set $S_2$ of states is locally distinguishable.
	
	ii) If $2m\equiv 0 \text{ mod } d$, then we must have $d=2m$. Hence we have the following transformations
	$$\{I, X^m, Z^n\} \xlongrightarrow{H_{2}}\{I, Z^{m}, X^n\}\xlongrightarrow{H_{2} }\textit{S}_{2,2}=\{I, X^{m}, X^{2n}Z^{-n}\}. $$
	This case can be separated into the following three cases:
	\begin{enumerate}
		\item  $2n \neq d \text{ mod }d $ and $2n \neq m \text{ mod }d$, then by Lemma \ref{lemma} the  set $\textit{S}_{2,2}$ is distinguishable.
		\item $2n \equiv d \text{ mod }d $ , then $d=2n$, and $m=n$. We only need to check the case $\{I, X^{n}, Z^{n}\},$   where $ d=2n $. This case will be solved below as the exceptional case 1.
		\item $2n \equiv m \text{ mod }d $ with $d=2m$, $\Rightarrow$ $2n=3m,$ $\Rightarrow$  $n=3k,m=2k$ for some integer $k$. We need to consider our first set  $S=\{I, X^{3k}, Z^{2k}\},$ where $d=4k$. This case will be solved below as the exceptional case 2.
	\end{enumerate}

	iii) If $2m\equiv n \text{ mod } d$, under the following transformation:
\begin{equation}
\{I, X^m, Z^n\} \xlongrightarrow{H_{4}}\textit{S}_{4}=\{I,X^{4m}Z^{-m}, X^n\},
\end{equation}
	clearly, we have $4m\neq n \text{ mod } d$. Then if $4m\neq 0\text{ mod } d$, by Lemma \ref{lemma}, we get the conclusion. And the case $4m \equiv 0\text{ mod } d$ imples that $ 2n\equiv d \equiv 0\text{ mod } d$, hence $d=2n$. Hence we have $2m\equiv n \text{ mod } 2n$ $\Rightarrow$ $2m=3n,$ $\Rightarrow$  $n=2k,m=3k$ for some integer $k$.  Then we only need to consider the case  $S_4= \{I, Z^{2k}, X^{3k}\}$ with $d=4k$.  This case will be solved below as the exceptional case 2.
	
Now we give an explicit strategies for Alice and Bob in order to distinguish the two sets of exceptional cases.
	
\textbf{\textit{Exceptional case 1:}}	
the case $\{I, X^{n}, Z^{n}\}$ with $d=2n$, $n\geq2$. 	
	The corresponding (unnormalized) states are shown below,
\begin{equation}
	\begin{array}{l}
	|\psi_1\rangle=|0, 0\rangle+|1, 1\rangle+|2,2\rangle+|3,3\rangle+...+|2n-2,2n-2\rangle+|2n-1,2n-1\rangle,\\
	|\psi_2\rangle=|n,0\rangle+|n+1,1\rangle+...+|2n-1, n-1\rangle+ |0,n\rangle+...+|n-1,2n-1\rangle,\\
	|\psi_3\rangle=|0,0\rangle-|1,1\rangle+|2,2\rangle-|3,3\rangle+...+|2n-2,2n-2\rangle-|2n-1,2n-1\rangle.
	\end{array}
\end{equation}
	Alice employ the following projective measurements:
	$M_k^{\pm}=(|2k-2\rangle\pm|2k-1\rangle)(\langle2k-2|\pm\langle2k-1|)$, $k=1,2,...,n$.
	The corresponding resulting states are, respectively,
\begin{equation}
	\begin{array}{l}
	|\widetilde{\psi_1}\rangle=(|2k-2\rangle\pm |2k-1\rangle)(|2k-2\rangle\pm |2k-1\rangle),\\
	|\widetilde{\psi_2}\rangle=(|2k-2\rangle\pm |2k-1\rangle)(|2k-2+n\rangle\pm |2k-1+n\rangle),\\
	|\widetilde{\psi_3}\rangle=(|2k-2\rangle\pm |2k-1\rangle)(|2k-2\rangle\mp|2k-1\rangle).
	\end{array}
\end{equation}
	Hence the states of Bob's system are orthogonal each other and Bob can distinguish the above three states $\{|\widetilde{\psi_1}\rangle,|\widetilde{\psi_2}\rangle,|\widetilde{\psi_3}\rangle\}$ exactly.
		
	\textbf{\textit{Exceptional case 2:}}
	  the case $\{I, X^{3k}, Z^{2k} \}$ with $d=4k$.
	  	The corresponding states are given below,
\begin{equation}
	\begin{array}{l}
	|\psi_1\rangle=|0, 0\rangle+|1, 1\rangle+|2,2\rangle+|3,3\rangle+...+|4k-2,4k-2\rangle+|4k-1,4k-1\rangle,\\
	|\psi_2\rangle=|3k,0\rangle+|3k+1,1\rangle+...+|4k-1,k-1\rangle+ |0,k \rangle+...+| 3k-1,4k-1\rangle,\\
	|\psi_3\rangle=|0,0\rangle-|1,1\rangle+|2,2\rangle-|3,3\rangle+...+|4k-2,4k-2\rangle-|4k-1,4k-1\rangle.
	\end{array}
\end{equation}
	Alice applies the following projective measurements:
	$M_l^{\pm}=(|2l-2\rangle\pm|2l-1\rangle)(\langle2l-2|\pm\langle2l-1|)$, $l=1,2,3,...,2k$.
correspondingly one gets
\begin{equation}
	\begin{array}{l}
	|\widetilde{\psi_1}\rangle=(|2l-2\rangle\pm |2l-1\rangle)(|2l-2\rangle\pm |2l-1\rangle),\\
	|\widetilde{\psi_2}\rangle=(|2l-2\rangle\pm |2l-1\rangle)(|2l-2+k\rangle\pm |2l-1+k\rangle),\\
	|\widetilde{\psi_3}\rangle=(|2l-2\rangle\pm |2l-1\rangle)(|2l-2\rangle\mp|2l-1\rangle).
	\end{array}
\end{equation}
	If $k\geq 2$, the states of Bob's system are orthogonal each other, and the states $\{|\widetilde{\psi_1}\rangle,|\widetilde{\psi_2}\rangle,|\widetilde{\psi_3}\rangle\}$ can be distinguished exactly. The case $k=1$ is   considered in Theorem 2 of the ref.\cite{Singal15} and was proved to be locally distinguished.
	\qed
		
	 The results of the above theorem can be understood as  a little step towards the  generalization of H. Fan's results in \cite{Fan04} to  arbitrary dimensional case.  Unlike the prime dimensional cases, sometimes, it may need to do several transformations before one could use Lemma \ref{lemma}. Moreover, one may encounter some exceptional cases which could not be dealt with by applying  Lemma \ref{lemma}.
	
	 The above results can be also understoood as part of results
	  toward the problem of locally distinguishability for any three orthgonal maximally entangled states. The results we obtained and those in \cite{Nathanson2013} give an evidence of positive answer. In \cite{Nathanson2013}, the authors presented some triple sets of maximally entangled states which are shown to be two-way  distinguishable by giving the explicit strategies. In our paper, we mainly devote to the set of generalized Bell states satisfying some conditions. Under these conditions we can transform complicated cases into some simple ones. However, there are also some exceptional cases for which the explicit constructions of strategies are needed.

	\section{Conclusion and discussion}
	
	In this paper, we have studied the problem of local distinguishability of   maximally entangled states, the generalized Bell states.  Firstly, we generalized some equations which have been considered by H. Fan for prime dimensional case to the case of arbitary dimensional ones. Since the local distinguishability of a set of quantum states is unchanged under local  unitary operations, we apply some local unitary operations to simplify the locally distinguished strategies.    By using the generalized equations   and giving the explicit strategies for some exceptional cases, we have obtained that any three generalized Bell states in $\mathbb{C}^d\otimes\mathbb{C}^d (d\geq4)$ are locally distinguishable.  However, the local distinguishablity of any three maximally entangled states in $\mathbb{C}^d\otimes\mathbb{C}^d (d\geq4)$ remains open.
	
	It is natural to ask whether four, five or more generalized Bell states can always be locally distinguished for large dimension $d$.  It seems that from what we have done in this paper, the case of four, five or more states can be similarly dealt with case by case.  However, the problem becomes more complicated.  Hence it is also interesting to develop other methods to solve these problems.
	
	\vspace{2.5ex}
	\noindent{\bf Acknowledgments}\, \, The authors thank the  referees for many helpful suggestions.
	This work is supported by the NSFC 11475178, NSFC 11571119 and NSFC 11675113.

\end{document}